\documentclass[
amsmath,amssymb,
pra,
]
{revtex4-2}

\usepackage{graphicx}
\usepackage{dcolumn}
\usepackage{bm}
\usepackage{hyperref}

\begin{document}

\title{ Radiation of Charge Moving along Face of Inverted Prism}

\author{Andrey V. Tyukhtin}
\email{a.tyuhtin@spbu.ru}
\affiliation{%
Saint Petersburg State University, 7/9 Universitetskaya nab., St. Petersburg, 199034 Russia
}%
\author{Sergey N. Galyamin}
\email{s.galyamin@spbu.ru}
\affiliation{%
Saint Petersburg State University, 7/9 Universitetskaya nab., St. Petersburg, 199034 Russia
}%
\author{Viktor V. Vorobev}%
 \email{vorobev@in.tum.de}
\affiliation{%
Technical University of Munich, Germany, Department of Informatics
}%

\maketitle

Cherenkov radiation (CR) generated by a charge moving along one of the faces of a dielectric prism is analyzed. 
Unlike our previous papers, here we suppose that the charge moves from the the prism ``nose'' to its base. 
For CR analysis, we use the technique described in our previous papers and called the ``aperture method''. 
However, here we develop a new version of this technique which is suitable for objects with plane faces: it utilizes field expansion only over plane waves inside the object. This approach is especially convenient for objects having two or more plane interfaces on which the waves are reflected and/or refracted. 
Using this technique, we obtain the electromagnetic field distribution over the aperture and then apply Stratton-Chu formulas (aperture integrals). Further, the main attention is paid to the calculation of the radiation field in the Fraunhofer (far-field) area. It is notable that we obtain expressions for corresponding Fourier transforms in the form of single integrals. Using them, the series of typical angular diagrams are computed and physical conclusions are made.

\section{Introduction}

Over the years, radiation of charged particles moving in presence of dielectric objects (``targets'' or ``radiators'') of complicated form was of essential interests for various applications. 
First, one can mention extensive studies of Cherenkov radiation (CR), both theoretical and experimental, connected with the development of Cherenkov detectors and counters~\cite{Jb, Zrb}.
Typical forms of used targets were cylinder, cone, sphere, prism or their combination~\cite{Getting47, Dicke47, Marshall51, Marshall52}.
Moreover, CR-based microwave radiation sources which utilize conical or prismatic radiators were also discussed at that time~\cite{Danos53, Coleman60} while the idea itself was introduced much earlier~\cite{G47}.
However, in recent decades, the aforementioned topics are again in the focus of scientific research due to several prospective applications of CR produced by high-quality beams of modern accelerators such as non-invasive bunch diagnostics systems and beam-driven radiation sources.
For example, experimental results on microwave and Terahertz CR generation from prismatic or conical targets have been reported recently~\cite{Takahashi00, Sei15, Sei17, KarPot2020}.
A prismatic radiator was proposed for non-invasive CR-based bunch diagnostics in~\cite{Pot10, PPSNS10}, this idea was further developed in succeeding papers~\cite{Bergamproc17, Kieffer18, Davut2021ipac, Lasocha2021ipac}.
Also a scheme of a beam-position monitor based on an interaction between a charged particle bunch and a pair of dielectric cylinders was actively developed in a series of recent publications~\cite{YevtushenkoNosich2019, Yevtushenko2020, HerasymovaNosich2021}.
   
Typically, the size of the aforementioned targets is much larger than the wavelengths of interest. 
On the one hand, this fact complicates considerably numerical simulations of the problem because they require a large amount of computational resources. 
However, on the other hand, this fact gives us a small parameter of the problem and allows the development of efficient approximate methods of analysis.
During last decade, we developed two such methods which can be called ``ray-optic method'' and ``aperture method''. 
Both these methods are valid for discussed objects (cylinders, cones, spheres, prisms) if their sizes are much larger than the wavelengths under consideration. 
Fundamentals of these methods and solution of different problems with the use of them are described in a series of our papers (most recent of them are the papers~\cite{TGV19, TVGB19, GVT19, GVT19E, TBGV20, TGVGrig20, TGV21_JOSAB, TGV22}). 
Note that the aperture method is more general than the ray optics one since the latter cannot describe diffraction effects and requires additional limitation on the observation point position).
Therefore we will use the aperture method throughout the paper.
It should be also emphasized that this approach has been benchmarked for various objects, in particular, dielectric cone and a ball with a vacuum channel (see~\cite{TGV19,TBGV20} and references therein). 
This benchmarking was performed by comparison with COMSOL simulation. 
As a result, it has been shown (as expected) that the aperture method gives an accuracy of the order of the ratio between the considered wavelength and the size of the object (at least in the regions of the maximal field magnitudes).

Earlier we have studied the charge radiation in the presence of a prismatic object however only in the case when the charge moves from the base of the prism to its top (the case of the ``direct'' prism)~\cite{TVGB19}. 
Here we consider the case when the charge moves from the top of the prism to its base (the case of ``inverted'' prism). 
This problem is much more complicated than the case of ``direct'' prism. 
The reason for this is that we have to take into account the reflection of radiation from the inclined face of the prism. 
The subsequent exit of this reflected radiation from the prism base can make the main contribution to the radiation field in the region outside the target. 

The radiation passing through such a path inside the prism can be calculated, for example, using the ray optics (taking into account, in particular, the divergence of the ray tube). 
However such analytical  calculation is  very difficult because we deal not only with the refraction on the base but also with the preliminary reflection from the oblique face. 

To avoid the complex analytical calculation of the field inside the object we will apply here a new version of the aperture method.
We use the fact that the target has only plane borders.
Therefore, unlike our previous works, inside the prism we represent the field only as an expansion in plane waves  (without calculating these integral representations).
The entire calculation of the integrals will concern only the area outside the prism.

\section{Aperture integrals: general form and approximation for Fraunhofer zone}

First of all, we remind the Straton-Chu formulas (known as well as ``aperture integrals'') in a form convenient for our researches~\cite{TGV19,TVGB19}. According to them the Fourier transform of electric field can be written in the following general form (we use Gaussian system of units):
\begin{equation}\label{eq:1.1}
    \begin{aligned}
    & \vec{E}\left( {\vec{R}} \right)={{{\vec{E}}}^{(h)}}\left( {\vec{R}} \right)+{{{\vec{E}}}^{(e)}}\left( {\vec{R}} \right), \\
    & {{{\vec{E}}}^{(h)}}\left( {\vec{R}} \right)=\frac{ik}{4\pi }\int\limits_{\Sigma }{\left\{ \left[ \vec{n}'\times \vec{H}\left( \vec{R}' \right) \right]G\left( \left| \vec{R}-\vec{R}' \right| \right)+ \right.} \\
    & \left. \,\,\,\,\,\,\,\,\,\,\,\,\,\,\,+\frac{1}{{{k}^{2}}}\left( \left[ \vec{n}'\times \vec{H}\left( \vec{R}' \right) \right]\cdot \nabla ' \right)\nabla 'G\left( \left| \vec{R}-\vec{R}' \right| \right) \right\}d\Sigma ', \\
    & {{{\vec{E}}}^{(e)}}\left( {\vec{R}} \right)=\frac{1}{4\pi }\int\limits_{\Sigma }{\left[ \left[ \vec{n}'\times \vec{E}\left( \vec{R}' \right) \right]\times \nabla 'G\left( \left| \vec{R}-\vec{R}' \right| \right) \right]d\Sigma ',} \\
    \end{aligned}
\end{equation}
where $\Sigma $ is an aperture area, $\vec{E}\left( \vec{R}' \right)$, $\vec{H}\left( \vec{R}' \right)$ is the field on the aperture, $k={\omega }/{c}$ is a wave number of the outer space (which is vacuum), $\vec{n}'$ is a unit external normal to the aperture in the point $\vec{R}'$ (it is directed into the area where the observation point is placed), $G\left( R \right)={\exp \left( i\,kR \right)}/{R}$ is a Green function of Helmholtz equation, and ${\nabla }'$ is a gradient: $\nabla '={{\vec{e}}_{x}}{\partial }/{\partial x'}+{{\vec{e}}_{y}}{\partial }/{\partial y'}+{{\vec{e}}_{z}}{\partial }/{\partial z'}$. Analogous formulas are known for the magnetic field as well, however we do not write them here because we are interested in a ``wave'' zone $kL\gg 1$ ($L$ is a distance from the aperture to the observation point) where $\left| {\vec{E}} \right|\approx \left| {\vec{H}} \right|$.

The observation point is often located far from the target, in other words, in the region where so-called ``wave parameter'' $D$ is large:
\begin{equation}\label{eq:1.2}
D\sim \lambda L/\Sigma \sim {\lambda L}/{{{d}^{2}}}\gg 1,
\end{equation}
where $\lambda ={2\pi }/{k}$ is a wavelength under consideration, and $\Sigma \sim {{d}^{2}}$ is a square of an aperture (we assume that the origin of the coordinate frame is located in the vicinity of the target; in this case $L\sim R$). This region (which usually called the Fraunhofer area, or far-field area) is the main interest for the present work. The condition~\eqref{eq:1.2} automatically results in the inequality
\begin{equation}\label{eq:1.3}
    R\gg d\cdot \left( {d}/{\lambda }\; \right)\gg d,
\end{equation}
because $d\gg \lambda $ in the problem under consideration. Using the inequalities~\eqref{eq:1.2},~\eqref{eq:1.3} one can  obtain the following approximate formulas for Fraunhofer area:
\begin{equation}\label{eq:1.4}
\begin{aligned}
  & {{{\vec{E}}}^{(h)}}\left( {\vec{R}} \right)\approx \frac{ik\exp \left( ikR \right)}{4\pi R}\int\limits_{\Sigma }{\left\{ \left[ \vec{n}'\times \vec{H}\left( \vec{R}' \right) \right]-{{{\vec{e}}}_{R}}\left( {{{\vec{e}}}_{R}}\cdot \left[ \vec{n}'\times \vec{H}\left( \vec{R}' \right) \right] \right) \right\}\exp \left( -ik{{{\vec{e}}}_{R}}\vec{R}' \right)d\Sigma '}, \\
 & {{{\vec{E}}}^{(e)}}\left( {\vec{R}} \right)\approx \frac{ik\exp \left( ikR \right)}{4\pi R}\int\limits_{\Sigma }{\left[ {{{\vec{e}}}_{R}}\times \left[ \vec{n}'\times \vec{E}\left( \vec{R}' \right) \right] \right]\exp \left( -ik{{{\vec{e}}}_{R}}\vec{R}' \right)d\Sigma ',} \\
\end{aligned}
\end{equation}
where ${{\vec{e}}_{R}}={{\vec{R}}}/{R}\;$.

\section{Solution of the ``key'' problem in the form of expansion on plane waves}
\begin{figure}
\centering
\includegraphics[width=8cm]{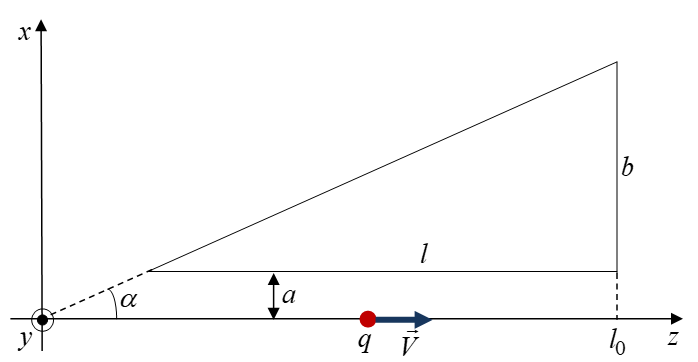}
\caption{\label{fig:1}%
The cross section of the prism.
}
\end{figure}

The geometry of the problem under consideration is shown in Fig.~\ref{fig:1}. The point charge $q$ moves along the ``lower'' face of the prism with velocity $\vec{V}=c\beta {{\vec{e}}_{z}}$. The z-axis is oriented on the charge velocity, and the x-axis is perpendicular to the lower face of the prism. The prism has the apex angle $\alpha $ and occupies the volume 
$x\in \left[ a, b+a \right]$, 
$y\in \left[ -{d}/{2}\;,{d}/{2}\; \right]$, 
$z\in \left[ a / \tan \alpha , l_0  \right]$, 
where $l_0=l+a / \tan \alpha $,    $l=b / \tan \alpha $. 
The prism material being isotropic and homogeneous are characterized by permittivity $\varepsilon $, permeability $\mu $, and the refractive index $n=\sqrt{\varepsilon \mu }$. It is assumed that these values do not depend on the wave vector (the spatial dispersion is absent) but they can be depend on the frequency $\omega $ (the frequency dispersion can be taken into account). The medium surrounding the prism is a vacuum.

The charge and current densities are determined correspondingly by the expressions $\rho =q\,\delta (x)\,\delta (y)\,\delta (z-Vt)$, $\vec{j}=V\rho \,{{\vec{e}}_{z}}$, where $\delta (\xi )$ is the Dirac delta-function. Note that we consider a point charge for definiteness only. The results will be obtained for the Fourier images of the field components, and they can be easily generalized to the case of a particle beam elongated along the z-axis.

Note that all expressions given further are expressions for the Fourier transforms ($\vec{E}$ and $\vec{H}$) of the field components. The components themselves are determined by Fourier integrals in the form $\int\limits_{-\infty }^{\infty }{\vec{E}{{e}^{-i\omega t}}d\omega }$. We will write out formulas for positive frequencies only ($\omega >0$). Expressions for negative frequencies are easily obtained by the rule $\vec{E}(-\omega )={{\vec{E}}^{*}}(\omega )$ (asterisk means complex conjugation) that follows from the reality of the field components.

The method applied by us uses the solution of certain ``key'' problem~\cite{TGV19,TVGB19,GVT19,GVT19E,TBGV20,TGVGrig20,TGVG20_PRA,TGV21_JOSAB}. In the case under consideration, it is the problem about the field of the charge moving along the boundary $x=a$ of the dielectric half-space $x>a$ and the vacuum half-space $x<a$. The solution of this problem is well-known~\cite{B62}; it was used by us for analysis of the problem with the ``direct'' prism~\cite{TVGB19}. However now we need to know the solution in the form of expansion on plane waves.

The field in vacuum half-space is presented as a sum of a field in infinite vacuum and a reflected field. Each of them consists of two polarizations and is presented as expansion in plane waves. The field in medium is an expansion in plane waves of two polarizations propagating from the boundary. Two reflection coefficients and two transmission ones are obtained from the boundary conditions (continuity of the tangent components of vectors $\vec{E}$ and $\vec{H}$). As a result of cumbersome but routine analytical calculations, one can obtain the following expressions for the field components in the medium (in area $x>a$):
\begin{equation}\label{eq:2.1}
{{\vec{E}}^{i}}=\int\limits_{-\infty }^{\infty }{\vec{E}_{0}^{i}{{e}^{iSx+i{{k}_{y}}y+{ikz}/{\beta }\;}}d{{k}_{y}}},\,\,\,\,\,\,{{\vec{H}}^{i}}=\int\limits_{-\infty }^{\infty }{\vec{H}_{0}^{i}{{e}^{iSx+i{{k}_{y}}y+{ikz}/{\beta }\;}}d{{k}_{y}}},
\end{equation}
\begin{equation}\label{eq:2.2}
\left\{ \begin{aligned}
  & E_{0x}^{i} \\
 & E_{0y}^{i} \\
 & E_{0z}^{i} \\
 & H_{0x}^{i} \\
 & H_{0y}^{i} \\
 & H_{0z}^{i} \\
\end{aligned} \right\}=\frac{q}{2\pi c}\left\{ \begin{aligned}
  & {{{T}_{1}}}/{(\beta \varepsilon )}\;+{{{T}_{2}}{{k}_{y}}}/{S}\; \\
 & {{{T}_{1}}{{k}_{y}}}/{(\beta \varepsilon S)}\;-{{T}_{2}} \\
 & {-{{T}_{1}}\left( {{n}^{2}}{{\beta }^{2}}-1 \right)k}/{({{\beta }^{2}}\varepsilon S)}\; \\
 & {-{{T}_{1}}{{k}_{y}}}/{S}+{{{T}_{2}}}/{(\beta \mu )}\; \\
 & {{T}_{1}}+{{{T}_{2}}{{k}_{y}}}/{(\beta \mu S)}\; \\
 & {-{{T}_{2}}\left( {{n}^{2}}{{\beta }^{2}}-1 \right)k}/{({{\beta }^{2}}\mu S)}\; \\
\end{aligned} \right\},
\end{equation}
where
\begin{equation}\label{eq:2.3}
\begin{aligned}
  & {{T}_{1}}=2\varepsilon S\frac{\mu \left( 1-{{\beta }^{2}} \right)S-i\left( {{n}^{2}}{{\beta }^{2}}-1 \right)K}{\Delta }{{e}^{-Ka-iSa}},\,\,\,\,\,\,\,{{T}_{2}}=\frac{2\mu \beta ({{n}^{2}}-1){{k}_{y}}S}{\Delta }{{e}^{-Ka-iSa}}; \\
 & \Delta =\left( {{n}^{2}}{{\beta }^{2}}-1 \right)\left[ \left( {{n}^{2}}-2{{n}^{2}}{{\beta }^{2}}+1 \right){{k}^{2}}{{\beta }^{-2}}+\left( {{n}^{2}}+1 \right)k_{y}^{2}-i\left( \varepsilon +\mu  \right)KS \right]; \\
 & S=\sqrt{{{s}^{2}}-k_{y}^{2}};\,\,\,\,\,\,\,\,\,\,
s=k \sqrt{{n^2} - \beta^{-2}};
\,\,\,\,\,\,\,\,\,\,K=\sqrt{{{k}^{2}}\left( {{\beta }^{-2}}-1 \right)+k_{y}^{2}};\,\,\,\,\,k={\omega }/{c}\;. \\
\end{aligned}
\end{equation}
The radicals are determined by the following rules: $\operatorname{Re}K>0$ (always), and $\operatorname{Im}S>0$ in the case of the losses in the medium taken into account. The last one gives that $\operatorname{sign} S=\operatorname{sign} \omega $ for propagating waves in the usual (no ``left-handed’’) medium. As we see the plane waves in integrands in~\eqref{eq:2.1} have the wave the wave vector
\begin{equation}\label{eq:2.4}
{{\vec{k}}^{i}}=S{{\vec{e}}_{x}}+{{k}_{y}}{{\vec{e}}_{y}}+k{{\beta }^{-1}}{{\vec{e}}_{z}},\,\,\,\,\,\,\,\,\,\,\,\left| {{{\vec{k}}}^{i}} \right|=kn,
\end{equation}
which makes the ``Cherenkov angle'' $\theta_p$ with the charge velocity:
\begin{equation}\label{eq:2.5}
{{\theta }_{p}}=\arccos \left( {1}/{\left( n\beta  \right)}\; \right).
\end{equation}

Formulas written above give the exact solution of the key problem. However the evanescent waves (having the imaginary values $S$) give exponentially small contribution in the field on relatively large distance $x$. Therefore we can replace integrals over the real axis with integrals over the finite interval:  $\int\limits_{-\infty }^{\infty }{...d{{k}_{y}}}\approx \int\limits_{-s}^{s}{...d{{k}_{y}}}$.

\section{Field of the wave incident directly on the base: the value at aperture}

\begin{figure}
\centering
\includegraphics[width=8cm]{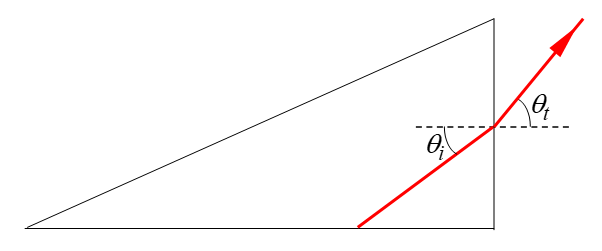}
\caption{\label{fig:2}%
The wave incident directly on the base.
}
\end{figure}
First, we consider the wave incident directly on the base of the prism which, for brevity, we will name the 1st wave (Fig.~\ref{fig:2}). The plane of incidence is formed by the normal to the aperture ${{\vec{e}}_{z}}$ and the wave vector of incident wave ${{\vec{k}}^{i}}$. The angles of incidence and refraction are, correspondingly,
\begin{equation}\label{eq:3.1}
{{\theta }_{i}}={{\theta }_{p}}=\arccos \left( {1}/{\left( n\beta  \right)}\; \right),\,\,\,\,\,\,\,\,{{\theta }_{t}}=\arcsin \left( n\sin {{\theta }_{i}} \right).
\end{equation}
Normal to the plane of incidence is
\begin{equation}\label{eq:3.2}
{{\vec{e}}_{\eta }}=\frac{\left[ {{{\vec{e}}}_{z}}\times {{{\vec{k}}}^{i}} \right]}{\left| \left[ {{{\vec{e}}}_{z}}\times {{{\vec{k}}}^{i}} \right] \right|}=\frac{-{{k}_{y}}\,{{{\vec{e}}}_{x}}+S\,{{{\vec{e}}}_{y}}}{s}.
\end{equation}

Let split the wave field into horizontal (the subscript $h$) and vertical (the subscript $v$ ) polarizations. For horizontal polarization, the electric field is orthogonal to the incidence plane, therefore for this wave $E_{0\eta }^{i}=\left( \vec{E}_{0}^{i}\cdot {{{\vec{e}}}_{\eta }} \right)$ 
(the subscript ``0'' everywhere denotes the amplitudes of the corresponding field components). 
For vertical polarization, the magnetic field is orthogonal to the incidence plane, therefore for this wave $H_{0\eta }^{i}=\left( \vec{H}_{0}^{i}\cdot {{{\vec{e}}}_{\eta }} \right)$. The magnetic field for horizontal polarization and electric field for vertical polarization can be fined with use of the rule that the electric field, the magnetic field and the wave vector are form right-handed orthogonal triplet (as well as we take into account that ${E}/{H}\;=\sqrt{{\mu }/{\varepsilon }\;}$). 
We obtain for horizontal polarization
\begin{equation}\label{eq:3.3}
\vec{E}_{h 0}^{i}
=E_{0\eta }^{i}{{\vec{e}}_{\eta }}
=-\frac{q}{2\pi c}\frac{s{{T}_{2}}}{S}{{\vec{e}}_{\eta }},\,\,\,\,\,\,\,\,\,\
\vec{H}_{h 0}^{i}
=\sqrt{\frac{\varepsilon }{\mu }}\left[ \frac{{{{\vec{k}}}^{i}}}{kn}\times {{{\vec{e}}}_{\eta }} \right]E_{0\eta }^{i},
\end{equation}
and for vertical polarization
\begin{equation}\label{eq:3.4}
\vec{H}_{v0}^{i}
=H_{0\eta }^{i}{{\vec{e}}_{\eta }}
=\frac{q}{2\pi c}\frac{s{{T}_{1}}}{S}{{\vec{e}}_{\eta }},\,\,\,\,\,\,\,\,
\vec{E}_{v0}^{i}
=-\sqrt{\frac{\mu }{\varepsilon }}\left[ \frac{{{{\vec{k}}}^{i}}}{kn}\times {{{\vec{e}}}_{\eta }} \right]H_{0\eta }^{i}.
\end{equation}
The wave vector of refracted wave is
\begin{equation}\label{eq:3.5}
{{\vec{k}}^{t}}=S\,{{\vec{e}}_{x}}+{{k}_{y}}\,{{\vec{e}}_{y}}+k\sqrt{1+{{\beta }^{-2}}-{{n}^{2}}}\,{{\vec{e}}_{z}}.
\end{equation}
Taking into account the refraction coefficients
\begin{equation}\label{eq:3.6}
{{T}_{h}}=\frac{2\cos {{\theta }_{i}}}{\cos {{\theta }_{i}}+\sqrt{{\mu }/{\varepsilon }\;}\cos {{\theta }_{t}}},\,\,\,\,\,\,\,\,{{T}_{v}}=\frac{2\sqrt{{\mu }/{\varepsilon }\;}\cos {{\theta }_{i}}}{\sqrt{{\mu }/{\varepsilon }\;}\cos {{\theta }_{i}}+\cos {{\theta }_{t}}}
\end{equation}
we obtain for horizontal polarization
\begin{equation}\label{eq:3.7}
\begin{aligned}
  & \vec{E}_{h0}^{t} = {{T}_{h}} E_{0\eta }^{i}{{{\vec{e}}}_{\eta }} = {{T}_{h}} E_{0\eta }^{i}\frac{-{{k}_{y}}\,{{{\vec{e}}}_{x}}+S\,{{{\vec{e}}}_{y}}}{s}, \\
 & \vec{H}_{h0}^{t}=  \left[ \frac{{{{\vec{k}}}^{t}}}{k}\times {{{\vec{e}}}_{\eta }} \right]E_{0\eta }^{t}=
 s^{-1}   \left[ -{{{\vec{e}}}_{x}} S \sqrt{1+{{\beta }^{-2}}-{{n}^{2}}} - {{{\vec{e}}}_{y}}{{k}_{y}}\sqrt{1+{{\beta }^{-2}}-{{n}^{2}}} + {{{\vec{e}}}_{z}} k \left( {{n}^{2}}-{{\beta }^{-2}} \right) \right] {{T}_{h}} E_{0\eta }^{i}, \\
\end{aligned}
\end{equation}
and for vertical polarization
\begin{equation}\label{eq:3.8}
\begin{aligned}
  & \vec{H}_{h0}^{t} = {{T}_{v}} H_{0\eta }^{i}{{{\vec{e}}}_{\eta }} = {{T}_{v}} H_{0\eta }^{i}\frac{-{{k}_{y}}\,{{{\vec{e}}}_{x}}+S\,{{{\vec{e}}}_{y}}}{s}, \\
 & \vec{E}_{v0}^{t}=-\left[ \frac{{{{\vec{k}}}^{t}}}{k}\times {{{\vec{e}}}_{\eta }} \right]H_{0\eta }^{t} = 
s^{-1} \left[ {{{\vec{e}}}_{x}}S\sqrt{1+{{\beta }^{-2}}- {{n}^{2}}} + {{{\vec{e}}}_{y}}{{k}_{y}}\sqrt{1+{{\beta }^{-2}}-{{n}^{2}}} - {{{\vec{e}}}_{z}}k\left( {{n}^{2}}-{{\beta }^{-2}} \right) \right] {{T}_{v}}  H_{0\eta }^{i}. \\
\end{aligned}
\end{equation}
Combining~\eqref{eq:3.7} and~\eqref{eq:3.8} we obtain for tangent components on the aperture in initial coordinates $x',\,y',\,z'=l_0$ the following expressions:
\begin{equation}\label{eq:3.9}
\{E_x, E_y, H_x, H_y \}=\int\limits_{-s}^{s} \{ E_{0x}, E_{0y}, H_{0x}, H_{0y}\} e^{iS{x}'+i{{k}_{y}}{y}'+i{k{{l}_{0}}}/{\beta }\;} d k_y ,
\end{equation}
\begin{equation}\label{eq:3.10}
\left\{ \begin{aligned}
  & E_{0x}^{{}} \\
 & E_{0y}^{{}} \\
 & H_{0x}^{{}} \\
 & H_{0y}^{{}} \\
\end{aligned} \right\}=\frac{q}{2\pi c S}\left\{ \begin{aligned}
 & {{T}_{2}}{{T}_{h}}{{k}_{y}} + {{T}_{1}}{{T}_{v}} S  \sqrt{1+{{\beta }^{-2}}-{{n}^{2}}} \\
 & -{{T}_{2}}{{T}_{h}}S + {{T}_{1}}{{T}_{v}} {{k}_{y}} \sqrt{1+{{\beta }^{-2}}-{{n}^{2}}} \\
 & -{{T}_{1}}{{T}_{v}}{{k}_{y}} + {{T}_{2}}{{T}_{h}} S \sqrt{1+{{\beta }^{-2}}-{{n}^{2}}} \\
 & {{T}_{1}}{{T}_{v}}S + {{T}_{2}}{{T}_{h}} {{k}_{y}} \sqrt{1+{{\beta }^{-2}}-{{n}^{2}}} \\
\end{aligned} \right\}.
\end{equation}

\section{Field of the wave reflected from the inclined face: the value at the aperture}

\begin{figure}
\centering
\includegraphics[width=8cm]{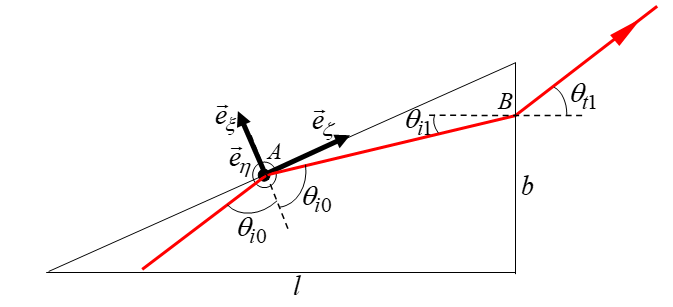}
\caption{\label{fig:3}%
The path of the wave reflected from the inclined face.
}
\end{figure}
Now let's consider the wave previously reflected from an inclined face (Fig.~\ref{fig:3}). For brevity, it will be named the 2nd wave (Fig.~\ref{fig:3}). This analysis is much more cumbersome due to the two interactions of the wave with two boundaries: reflection from an inclined face and refraction at the base.

The projection of the wave vector of incident wave makes the angle ${{\psi }_{i}}=\arccos \left( {k}/{\sqrt{{{k}^{2}}+{{\beta }^{2}}k_{y}^{2}}}\; \right)$ with the z-axis. We introduce the orts ${{\vec{e}}_{\xi }}$, ${{\vec{e}}_{\eta }}$, ${{\vec{e}}_{\zeta }}$ by the following way: ${{\vec{e}}_{\xi }}$ is the normal to the inclined face and ${{\vec{e}}_{\eta }}$ is the perpendicular to the plane of incidence:
\begin{equation}\label{eq:4.1}
\begin{aligned}
  & {{{\vec{e}}}_{\xi }}={{{\vec{e}}}_{x}}\cos \alpha -{{{\vec{e}}}_{z}}\sin \alpha , \\
 & {{{\vec{e}}}_{\eta }}=\frac{\left[ {{{\vec{k}}}^{i}}\times {{{\vec{e}}}_{\xi }} \right]}{\left| \left[ {{{\vec{k}}}^{i}}\times {{{\vec{e}}}_{\xi }} \right] \right|}=\frac{-{{k}_{y}}\sin \alpha \,{{{\vec{e}}}_{x}}+\left( S\sin \alpha +k{{\beta }^{-1}}\cos \alpha  \right){{{\vec{e}}}_{y}}-{{k}_{y}}\cos \alpha \,{{{\vec{e}}}_{z}}}{\sqrt{{{\left( S\sin \alpha +k{{\beta }^{-1}}\cos \alpha  \right)}^{2}}+k_{y}^{2}}}. \\
\end{aligned}
\end{equation}
(Note that some designations in this section (for example, $\eta $) coincide with the designations in the previous section, however they have a different meaning.) 
The normal ($\xi $) and tangential ($\tau $) components of the wave vector are, correspondingly,
\begin{equation}\label{eq:4.2}
k_{\xi }^{i}=S\cos \alpha - k \beta^{-1} \sin \alpha ,\,\,\,\,\,\,\,k_{\tau }^{i}=\sqrt{{{k}^{2}}{{n}^{2}}-{{\left( k_{\xi }^{i} \right)}^{2}}}.
\end{equation}

Let split the field into horizontal (the subsript $h$) and vertical (the subscript $v$) polarizations. For horizontal polarization, the amplitude of component orthogonal to the incidence plane is $E_{0\eta }^{i}=\left( \vec{E}_{0}^{i}\cdot {{{\vec{e}}}_{\eta }} \right)$ and, for vertical polarization, it is $H_{0\eta }^{i}=\left( \vec{H}_{0}^{i}\cdot {{{\vec{e}}}_{\eta }} \right)$:
\begin{equation}\label{eq:4.3}
\left\{ \begin{aligned}
  & E_{0\eta }^{i} \\
 & H_{0\eta }^{i} \\
\end{aligned} \right\}=\frac{q}{2\pi c}\left\{ \begin{aligned}
  & \mu  k k_y \cos \alpha \cdot {{T}_{1}}\, - \,\left[ {{s}^{2}}\sin \alpha 
+ k  \beta^{-1} S \cos \alpha  \right]\cdot {{T}_{2}} \\
 & \left[ {{s}^{2}}\sin \alpha +  k  \beta^{-1}  S \cos \alpha  \right]\cdot {{T}_{1}}
+ \varepsilon  k k_y \cos \alpha \cdot {{T}_{2}} \\
\end{aligned} \right\}\frac{1}{S\left| \left[ {{{\vec{k}}}^{i}}\times {{{\vec{e}}}_{\eta }} \right] \right|}.
\end{equation}
When the wave is reflected from an inclined face, the tangent component of the wave vector ${{\vec{k}}_{\tau }}$ does not change, and the normal one ${{k}_{\xi }}$ changes the sign, i.e. for reflected wave we have
\begin{equation}\label{eq:4.4}
k_{\xi }^{r0}=-k_{\xi }^{i}=-S\cos \alpha + k  \beta^{-1}  \sin \alpha ,\,\,\,\,\,\,\,k_{\tau }^{r0}=k_{\tau }^{i}=\sqrt{{{k}^{2}}{{n}^{2}}-{{\left( k_{\xi }^{i} \right)}^{2}}}.
\end{equation}
The angles of incidence and refraction on the oblique face are correspondingly
\begin{equation}\label{eq:4.5}
{{\theta }_{i0}}=\arcsin \left( {k_{\tau }^{i}}/{(kn)}\; \right),\,\,\,\,\,\,\,\,{{\theta }_{t0}}=\arcsin \left( n\sin {{\theta }_{i0}} \right)=\arcsin \left( {k_{\tau }^{i}}/{k}\; \right).
\end{equation}
Reflection coefficient for horizontal polarization is
\begin{equation}\label{eq:4.6}
	{{R}_{h0}}=\frac{\cos {{\theta }_{i}}-\sqrt{{\mu }/{\varepsilon }\;}\cos {{\theta }_{t0}}}{\cos {{\theta }_{i}}+\sqrt{{\mu }/{\varepsilon }\;}\cos {{\theta }_{t0}}}\,\,\,\,\,\,\,\text{or}\,\,\,\,\,\,{{R}_{h0}}=-1\,\,\,\text{for}\,\,\text{metalized}\,\,\text{face}\text{.}
\end{equation}
Reflection coefficient for vertical polarization is
\begin{equation}\label{eq:4.7}
	{{R}_{v0}}=\frac{\sqrt{{\mu }/{\varepsilon }\;}\cos {{\theta }_{i}}-\cos {{\theta }_{t0}}}{\sqrt{{\mu }/{\varepsilon }\;}\cos {{\theta }_{i}}+\cos {{\theta }_{t0}}}\,\,\,\,\,\,\,\text{or}\,\,\,\,\,\,{{R}_{v0}}=1\,\,\,\text{for}\,\,\text{metalized}\,\,\text{face}\text{.}
\end{equation}
Here, the first option refers to the case of the dielectric - vacuum interface, and the second option refers to the case of the metalized boundary.

It is convenient as well to write the wave vectors using coordinates $\tilde{\xi }=\xi ,\,\,\,\tilde{\eta }=y,\,\,\,\tilde{\zeta }$, where $\tilde{\zeta }$ is in the plane of incidence and parallel to the oblique face. We have:
\begin{equation}\label{eq:4.8}
 k_{{\tilde{\xi }}}^{i}=-k_{{\tilde{\xi }}}^{r0}=S\cos \alpha - 
k \beta^{-1}       \sin \alpha ,\,\,\,\,\,k_{{\tilde{\eta }}}^{i}=k_{{\tilde{\eta }}}^{r0}={{k}_{y}},\,\,\,\,\,k_{{\tilde{\zeta }}}^{i}=k_{{\tilde{\zeta }}}^{r0}=S\sin \alpha +  k \beta^{-1}  \cos \alpha .
\end{equation}
From here it is simple to find the components of ${{\vec{k}}^{r0}}$ in initial coordinates $x,\,y,\,z$:
\begin{equation}\label{eq:4.9}
	\begin{aligned}
  & k_{x}^{r0}=k_{\xi }^{r0}\cos \alpha +k_{{\tilde{\zeta }}}^{r0}\sin \alpha 
=   k \beta^{-1}    \sin \left( 2\alpha  \right)-S\cos \left( 2\alpha  \right), \\
 & k_{y}^{r0}={{k}_{y}}, \\
 & k_{z}^{r0}=-k_{\xi }^{r0}\sin \alpha +k_{{\tilde{\zeta }}}^{r0}\cos \alpha =k \beta^{-1} 
\cos \left( 2\alpha  \right)+S\sin \left( 2\alpha  \right) . \\
\end{aligned}
\end{equation}
Thus we obtain for the field reflected from the inclined face the following expression:
\begin{equation}\label{eq:4.10}
\left\{ \begin{aligned}
  & E_{\eta }^{r0} \\
 & H_{\eta }^{r0} \\
\end{aligned} \right\}=\int\limits_{-s}^{s}{\left\{ \begin{aligned}
  & {{R}_{h0}}E_{0\eta }^{i} \\
 & {{R}_{v0}}H_{0\eta }^{i} \\
\end{aligned} \right\}\exp \left\{ i\Phi (x,y,z) \right\}d{{k}_{y}}}\,,
\end{equation}
\begin{equation}\label{eq:4.11}
\Phi (x,y,z)= S\tilde{x}+{{k}_{y}}\tilde{y}+  k \beta^{-1}  \tilde{z}           ,\,+\,\,k_{x}^{r0}\left( x-\tilde{x} \right)+{{k}_{y}}\left( y-\tilde{y} \right)+k_{z}^{r0}\left( z-\tilde{z} \right),
\end{equation}
where $(\tilde{x},\tilde{y},\tilde{z})$ is the point where the wave is reflected. Since the waves are transversal we can find “additional” components by known formulas. We have for horizontal and vertical polarization, correspondingly,
\begin{equation}\label{eq:4.12}
\vec{H}_{0\,h}^{r0}=\sqrt{\frac{\varepsilon }{\mu }}\left[ \frac{{{{\vec{k}}}^{r0}}}{kn}\times {{{\vec{e}}}_{\eta }} \right]{{R}_{h0}}E_{0\eta }^{i},\,\,\,\,\,\,\,\vec{E}_{0\,v}^{r0}=-\sqrt{\frac{\mu }{\varepsilon }}\left[ \frac{{{{\vec{k}}}^{r0}}}{kn}\times {{{\vec{e}}}_{\eta }} \right]{{R}_{v0}}H_{0\eta }^{i0}.
\end{equation}
Components of  the cross product included here can be written in the following form:
\begin{equation}\label{eq:4.13}
\begin{aligned}
  & {{\left[ \frac{{{{\vec{k}}}^{r0}}}{kn}\times {{{\vec{e}}}_{\eta }} \right]}_{x}}=\frac{\left[ k{{\beta }^{-1}}\sin \left( 2\alpha  \right)-S\cos \left( 2\alpha  \right) \right]\left( k{{\beta }^{-1}}\sin \alpha -S\cos \alpha  \right)-{{k}^{2}}{{n}^{2}}\cos \alpha }{kn\left| \left[ {{{\vec{k}}}^{r0}}\times {{{\vec{e}}}_{\xi }} \right] \right|}, \\
 & {{\left[ \frac{{{{\vec{k}}}^{r0}}}{kn}\times {{{\vec{e}}}_{\eta }} \right]}_{y}}=\frac{\left( k{{\beta }^{-1}}\sin \alpha -S\cos \alpha  \right){{k}_{y}}}{kn\left| \left[ {{{\vec{k}}}^{r0}}\times {{{\vec{e}}}_{\xi }} \right] \right|}, \\
 & {{\left[ \frac{{{{\vec{k}}}^{r0}}}{kn}\times {{{\vec{e}}}_{\eta }} \right]}_{z}}=\frac{\left[ k{{\beta }^{-1}}\cos \left( 2\alpha  \right)+S\sin \left( 2\alpha  \right) \right]\left( k{{\beta }^{-1}}\sin \alpha -S\cos \alpha  \right)+{{k}^{2}}{{n}^{2}}\sin \alpha }{kn\left| \left[ {{{\vec{k}}}^{r0}}\times {{{\vec{e}}}_{\xi }} \right] \right|}. \\
\end{aligned}
\end{equation}

The field reflected from the oblique face is the field incident on the prism base. We need to know the components parallel to the base. Using initial coordinates one can obtain for the amplitudes of $x$- and $y$- components:
\begin{equation}\label{eq:4.14}
\begin{aligned}
  & E_{0\,x}^{r0}=-{{R}_{h0}}E_{0\eta }^{i}\frac{{{k}_{y}}\sin \alpha }{\left| \left[ {{{\vec{k}}}^{i}}\times {{{\vec{e}}}_{\xi }} \right] \right|}-{{R}_{v0}}H_{0\eta }^{i}\sqrt{\frac{\mu }{\varepsilon }}{{\left[ \frac{{{{\vec{k}}}^{r0}}}{kn}\times {{{\vec{e}}}_{\eta }} \right]}_{x}}, \\
 & E_{0y}^{r0}={{R}_{h0}}E_{0\eta }^{i}\frac{S\sin \alpha + k \beta^{-1} \cos \alpha }{\left| \left[ {{{\vec{k}}}^{i}}\times {{{\vec{e}}}_{\xi }} \right] \right|}-{{R}_{v0}}H_{0\eta }^{i}\sqrt{\frac{\mu }{\varepsilon }}{{\left[ \frac{{{{\vec{k}}}^{r0}}}{kn}\times {{{\vec{e}}}_{\eta }} \right]}_{y}}, \\
 & H_{0x}^{r0}=-{{R}_{v0}}H_{0\eta }^{i}\frac{{{k}_{y}}\sin \alpha }{\left| \left[ {{{\vec{k}}}^{i}}\times {{{\vec{e}}}_{\xi }} \right] \right|}+{{R}_{h0}}E_{0\eta }^{i}\sqrt{\frac{\varepsilon }{\mu }}{{\left[ \frac{{{{\vec{k}}}^{r0}}}{kn}\times {{{\vec{e}}}_{\eta }} \right]}_{x}}, \\
 & H_{0y}^{r0}={{R}_{v0}}H_{0\eta }^{i}\frac{S\sin \alpha +  k \beta^{-1}  \cos \alpha }{\left| \left[ {{{\vec{k}}}^{i}}\times {{{\vec{e}}}_{\xi }} \right] \right|}+{{R}_{h0}}E_{0\eta }^{i}\sqrt{\frac{\varepsilon }{\mu }}{{\left[ \frac{{{{\vec{k}}}^{r0}}}{kn}\times {{{\vec{e}}}_{\eta }} \right]}_{y}}. \\
\end{aligned}
\end{equation}

Since the normal to the base is ${{\vec{e}}_{z}}$ then the normal to the incident plane is
\begin{equation}\label{eq:4.15}
{{\vec{e}}_{\bot }}=\frac{\left[ {{{\vec{e}}}_{z}}\times {{{\vec{k}}}^{r0}} \right]}{\left| \left[ {{{\vec{e}}}_{z}}\times {{{\vec{k}}}^{r0}} \right] \right|}.
\end{equation}
Using~\eqref{eq:4.9} one can obtain
\begin{equation}\label{eq:4.16}
{{\vec{e}}_{\bot }}=\frac{-{{k}_{y}}{{{\vec{e}}}_{x}}+\left[   k \beta^{-1}  \sin \left( 2\alpha  \right)-S\cos \left( 2\alpha  \right) \right]{{{\vec{e}}}_{y}}}{\sqrt{k_{y}^{2}+{{\left[   k \beta^{-1}  \sin \left( 2\alpha  \right)-S\cos \left( 2\alpha  \right) \right]}^{2}}}}.
\end{equation}
The amplitudes of components orthogonal to the plane of incidence are
\begin{equation}\label{eq:4.17}
\begin{aligned}
  & E_{0\bot }^{r0}=\left( \vec{E}_{0}^{r0}\cdot {{{\vec{e}}}_{\bot }} \right)=\frac{-{{k}_{y}}E_{0\,x}^{r0}+\left[ k{{\beta }^{-1}}\sin \left( 2\alpha  \right)-S\cos \left( 2\alpha  \right) \right]E_{0\,y}^{r0}}{\sqrt{k_{y}^{2}+{{\left[ k{{\beta }^{-1}}\sin \left( 2\alpha  \right)-S\cos \left( 2\alpha  \right) \right]}^{2}}}}, \\
 & H_{0\bot }^{r0}=\left( \vec{H}_{0}^{r0}\cdot {{{\vec{e}}}_{\bot }} \right)=\frac{-{{k}_{y}}H_{0\,x}^{r0}+\left[ k{{\beta }^{-1}}\sin \left( 2\alpha  \right)-S\cos \left( 2\alpha  \right) \right]H_{0\,y}^{r0}}{\sqrt{k_{y}^{2}+{{\left[ k{{\beta }^{-1}}\sin \left( 2\alpha  \right)-S\cos \left( 2\alpha  \right) \right]}^{2}}}}. \\
\end{aligned}
\end{equation}
With respect to the base of the prism, the first of formulas~\eqref{eq:4.17} represents horizontal polarization, and the second one represents vertical polarization.

The wave vector of the refracted wave has the following components:
\begin{equation}\label{eq:4.18}
k_{x}^{t1}=k{{\beta }^{-1}}\sin \left( 2\alpha  \right)-S\cos \left( 2\alpha  \right),\,\,\,\,\,\,\,\,k_{y}^{t1}={{k}_{y}},\,\,\,\,\,\,\,k_{z}^{t1}=\sqrt{{{k}^{2}}-{{\left( k_{x}^{t1} \right)}^{2}}-k_{y}^{2}}\,.
\end{equation}
The angles of incidence and refraction are, correspondingly,
\begin{equation}\label{eq:4.19}
{{\theta }_{i1}}=\arccos \left( {k_{z}^{r0}}/{\left( kn \right)}\; \right),\,\,\,\,\,\,\,\,{{\theta }_{t1}}=\arcsin \left( n\sin {{\theta }_{i1}} \right).
\end{equation}
Coefficients of refraction for horizontal and vertical polarization are, correspondingly,
\begin{equation}\label{eq:4.20}
{{T}_{h1}}=\frac{2\cos {{\theta }_{i1}}}{\cos {{\theta }_{i1}}+\sqrt{{\mu }/{\varepsilon }\;}\cos {{\theta }_{t1}}},\,\,\,\,\,\,\,\,{{T}_{v1}}=\frac{2\sqrt{{\mu }/{\varepsilon }\;}\cos {{\theta }_{i1}}}{\sqrt{{\mu }/{\varepsilon }\;}\cos {{\theta }_{i1}}+\cos {{\theta }_{t1}}}.
\end{equation}
Amplitudes of base components of both polarizations are
\begin{equation}\label{eq:4.21}
E_{0\bot }^{t1}={{T}_{h1}}E_{0\bot }^{r0},\,\,\,\,\,\,\,H_{0\bot }^{t1}={{T}_{v1}}H_{0\bot }^{r0}.
\end{equation}
Additional components (parallel to the plane of refraction)
\begin{equation}\label{eq:4.22}
\vec{H}_{0\,\parallel }^{t1}=-E_{0\,\bot }^{t1}\left[ {{{\vec{e}}}_{\bot }}\times {{{{\vec{k}}}^{t1}}}/{k}\; \right],\,\,\,\,\,\,\,\,\,\,\vec{E}_{0\,\parallel }^{t1}=H_{0\,\bot }^{t1}\left[ {{{\vec{e}}}_{\bot }}\times {{{{\vec{k}}}^{t1}}}/{k}\; \right].
\end{equation}
Using~\eqref{eq:4.16} and~\eqref{eq:4.18} one can obtain $x$- and $y$-components of the fields on the aperture:
\begin{equation}\label{eq:4.23}
\left\{ \begin{aligned}
  & E_{0x}^{{}} \\
 & E_{0y}^{{}} \\
 & H_{0x}^{{}} \\
 & H_{0y}^{{}} \\
\end{aligned} \right\}=\frac{1}{k\,\sqrt{k_{y}^{2}+{{\left( k_{x}^{t1} \right)}^{2}}}}\left\{ \begin{aligned}
  & -k{{k}_{y}}\,E_{0\bot }^{t1}+k_{x}^{t1}k_{z}^{t1}\,H_{0\bot }^{t1} \\
 & kk_{x}^{t1}\,\,E_{0\bot }^{t1}+{{k}_{y}}k_{z}^{t1}\,H_{0\bot }^{t1} \\
 & -k{{k}_{y}}H_{0\bot }^{t1}-k_{x}^{t1}k_{z}^{t1}\,E_{0\bot }^{t1} \\
 & kk_{x}^{t1}\,H_{0\bot }^{t1}-{{k}_{y}}k_{z}^{t1}\,E_{0\bot }^{t1} \\
\end{aligned} \right\}.
\end{equation}

Let us now calculate the phase of the wave, which, according to~\eqref{eq:4.11}, can be written at an arbitrary point $B=\left( {x}',\,{y}',\,{z}'={{l}_{0}} \right)$ on the aperture in the form
\begin{equation}\label{eq:4.24}
\Phi ({x}',{y}',{{k}_{y}})=S\tilde{x} +   k \beta^{-1}   \tilde{z}     \,\,+\,\,k_{x}^{r0}\left( {x}'-\tilde{x} \right)+{{k}_{y}}{y}'+k_{z}^{r0}\left( {{l}_{0}}-\tilde{z} \right).
\end{equation}
Remind that $A=\left( \tilde{x},\,\tilde{y},\,\tilde{z} \right)$ is the point of reflection from the oblique face (Fig.~\ref{fig:2}). It has to be expressed through coordinates $\left( {x}',\,{y}',\,{z}'={{l}_{0}} \right)$.
The straight line $AB$ (Fig.~\ref{fig:2}) is determined by the equations	
\begin{equation}\label{eq:4.25}
\frac{x-{x}'}{k_{x}^{r0}}=\frac{y-{y}'}{{{k}_{y}}}=\frac{z-{{l}_{0}}}{k_{z}^{r0}}.
\end{equation}
The oblique face is determined by equation
\begin{equation}\label{eq:4.26}
x=z\tan \alpha.
\end{equation}
Solving the system~\eqref{eq:4.25},~\eqref{eq:4.26} and taken into account formulas~\eqref{eq:4.9} we obtain the coordinates of the point of reflection $A$:
\begin{equation}\label{eq:4.27}
\tilde{z}=\frac{k_{z}^{r0}{x}'-k_{x}^{r0}{{l}_{0}}}{S -   k \beta^{-1}   \tan \alpha },\,\,\,\,\,\,\tilde{x}=\tilde{z}\tan \alpha =\frac{k_{z}^{r0}{x}'-k_{x}^{r0}{{l}_{0}}}{S-  k \beta^{-1}  \tan \alpha }\tan \alpha ,\,\,\,\,\,\,\tilde{y}={y}'+{{k}_{y}}\frac{\tilde{z}-{{l}_{0}}}{k_{z}^{r0}}.\,
\end{equation}
The differences of coordinates of $A$ and $B$ are equal to the following:
\begin{equation}\label{eq:4.28}
\left\{ \begin{aligned}
  & {x}'-\tilde{x} \\
 & {y}'-\tilde{y} \\
 & {{l}_{0}}-\tilde{z} \\
\end{aligned} \right\}=\frac{{{l}_{0}}\tan \alpha -{x}'}{S-  k \beta^{-1}  \tan \alpha }\left\{ \begin{aligned}
  & k_{x}^{r0} \\
 & {{k}_{y}} \\
 & k_{z}^{r0} \\
\end{aligned} \right\}.
\end{equation}
Substituting~\eqref{eq:4.27},~\eqref{eq:4.28} in~\eqref{eq:4.24} after some transformation one can obtain the following expression for the phase on the aperture:
\begin{equation}\label{eq:4.29}
\begin{aligned}
  & \Phi ({x}',{y}',{{k}_{y}})={{\kappa }_{x}}{x}'+{{\kappa }_{y}}{y}'+{{\kappa }_{z}}{{l}_{0}}; \\
 & {{\kappa }_{x}} = \frac{1} {S-k{{\beta }^{-1}}\tan \alpha }
\left[ \frac{ {{k}^{2}}  \cos \left( 3\alpha  \right)}  {\beta^2 \cos \alpha } 
- \left( k^2 n^2 - k_{y}^{2} \right) \cos \left( 2\alpha  \right)
+ \frac{  k S   \sin \left( 3\alpha  \right)}{ \beta  \cos \alpha } \right], \\
 & {{\kappa }_{y}}={{k}_{y}}, \\
 & {{\kappa }_{z}}=\frac{1}{S-k{{\beta }^{-1}}\tan \alpha }
\left[ - \frac{ {{k}^{2}} \sin \left( 3\alpha  \right)}{ \beta^2  \cos \alpha } 
+ \left( k^2 n^2 - k_{y}^{2} \right) \sin \left( 2\alpha  \right) 
+ \frac{ kS     \cos \left( 3\alpha  \right)} { \beta \cos \alpha } \right]. \\
\end{aligned}
\end{equation}

Thus we have obtained the tangent components of the field in arbitrary point $\left( {x}',\,{y}',\,{z}'={{l}_{0}} \right)$ on the aperture in the following form:
\begin{equation}\label{eq:4.30}
\left\{ \begin{aligned}
  & {{E}_{x}}({x}',{y}',l_0)\, \\
 & {{E}_{y}}({x}',{y}',l_0)\, \\
 & {{H}_{x}}({x}',{y}',l_0)\, \\
 & {{H}_{y}}({x}',{y}',l_0) \\
\end{aligned} \right\}=\int\limits_{-s}^{s}{\left\{ \begin{aligned}
  & {{E}_{0x}}({{k}_{y}}) \\
 & {{E}_{0y}}({{k}_{y}})\, \\
 & {{H}_{0x}}({{k}_{y}})\, \\
 & {{H}_{0y}}({{k}_{y}}) \\
\end{aligned} \right\}{{e}^{i\Phi ({x}',{y}',{{k}_{y}})}}d{{k}_{y}}},
\end{equation}
where the phase $\Phi ({x}',{y}',{{k}_{y}})$ is determined be Eqs.~\eqref{eq:4.29} and the amplitudes ${{E}_{0x,y}}({{k}_{y}})$, ${{H}_{0x,y}}({{k}_{y}})$ are determined by formulas~\eqref{eq:4.23},~\eqref{eq:4.17},~\eqref{eq:4.14} and~\eqref{eq:4.3} (taking into account all the notations introduced above). Of course, the results seem very cumbersome, but they should not present large difficulties for computer calculations.

\section{Radiation in Fraunhofer area}

Now we can calculate the field in the area outside the target using the Stratton-Chu formulas. To calculate the field at an arbitrary point, it is needed to use the formulas~\eqref{eq:1.1}. However the most interesting is the Fraunhofer area in which the radiation directivity pattern is formed. This area is characterized by large wave parameter ($D\gg1$), and we can apply the asymptotic formulas~\eqref{eq:1.4}.

We will use traditional spherical coordinates $R,\,\theta ,\,\varphi $ connecting with Cartesian coordinates by relations
\begin{equation}\label{eq:5.1}
x=R\sin \theta \cos \varphi ,\,\,\,\,\,y= R\sin \theta \sin \varphi ,\,\,\,\,\,z=R\cos \theta .
\end{equation}
Taking into account that ${\vec{n}}'={{\vec{e}}_{z}}$, after calculation of projections of the integrands in~\eqref{eq:1.4} on the directions ${{\vec{e}}_{R}}$, ${{\vec{e}}_{\theta }}$, ${{\vec{e}}_{\varphi }}$ one can obtain the following expressions:
\begin{equation}\label{eq:5.2}
\begin{aligned}
  & \left\{ \begin{aligned}
  & {{E}_{R}}\left( {\vec{R}} \right) \\
 & {{E}_{\theta }}\left( {\vec{R}} \right) \\
 & {{E}_{\varphi }}\left( {\vec{R}} \right) \\
\end{aligned} \right\}\approx \frac{ik\exp \left( ikR \right)}{4\pi R}\times  \\
 & \times \int\limits_{\Sigma }{\begin{aligned}
  & \left\{ \begin{aligned}
  & 0 \\
 & -{{E}_{x}}\left( {{\vec{R}}'} \right)\cos \varphi -{{E}_{y}}\left( {{\vec{R}}'} \right)\sin \varphi +\left[ {{H}_{x}}\left( {{\vec{R}}'} \right)\sin \varphi -{{H}_{y}}\left( {{\vec{R}}'} \right)\cos \varphi  \right]\cos \theta  \\
 & \left[ {{E}_{x}}\left( {{\vec{R}}'} \right)\sin \varphi -{{E}_{y}}\left( {{\vec{R}}'} \right)\cos \varphi  \right]\cos \theta +{{H}_{x}}\left( {{\vec{R}}'} \right)\cos \varphi +{{H}_{y}}\left( {{\vec{R}}'} \right)\sin \varphi  \\
\end{aligned} \right\}\times  \\
 & \,\,\,\,\,\,\,\,\,\,\,\,\,\,\,\,\,\,\,\,\,\,\,\,\,\,\,\,\,\,\,\,\,\,\,\,\,\,\,\,\,\,\times \exp \left\{ -ik\left( {x}'\sin \theta \cos \varphi +{y}'\sin \theta \sin \varphi +{z}'\cos \theta  \right) \right\}d{\Sigma }' .\\
\end{aligned}} \\
\end{aligned}
\end{equation}
As we see ${{E}_{R}}=0$ that was expected because of the waves are transversal with respect to $\vec{R}$.

Remind that the aperture $\Sigma $ is the part of the prism base that is illuminated by radiation. The 1st wave always illuminates the entire base, but the 2nd wave can illuminate both the entire base and its part, depending on the problem parameters. The boundary of the illuminated area $x'=x'_b$ is determined by the wave reflected from the inclined face in the prism top. To find this value it should be substitute $\tilde{x}={{\tilde{z}}}/{\tan \alpha }\;=a$ in the 2nd equation of~\eqref{eq:4.27}. As result we obtain:
\begin{equation}\label{eq:5.3}
x'_b  =  \frac{a\left( S-k{{\beta }^{-1}}\tan \alpha  \right)+k_{x}^{r0}{{l}_{0}}\tan \alpha }{k_{z}^{r0}\tan \alpha }\approx \frac{k_{x}^{r0}}{k_{z}^{r0}}{{l}_{0}},
\end{equation}
If $  x'_b < a  $ then the lower boundary of the aperture is ${x}'=a$, but in opposite case it is $x'=x'_b$. Thus the the lower boundary of the region of integration with respect to ${x}'$ is
\begin{equation}\label{eq:5.4}
h=\max (a,   x'_b  )=\max \left( a,\,\,\frac{a\left( {S}/{\tan \alpha }\;-{k}/{\beta }\; \right)+k_{x}^{r0}{{l}_{0}}}{k_{z}^{r0}} \right).
\end{equation}
We emphasize that $h$ depends on $k_y$. 
Substituting~\eqref{eq:3.9} or~\eqref{eq:4.30} in~\eqref{eq:5.2} we obtain
\begin{equation}\label{eq:5.5}
\left\{ \begin{aligned}
  & {{E}_{\theta }}\left( {\vec{R}} \right) \\
 & {{E}_{\varphi }}\left( {\vec{R}} \right) \\
\end{aligned} \right\}   =
\frac{ik\exp \left( ikR \right)}{4\pi R}\int\limits_{-s}^{s}{d{{k}_{y}}\left\{ \begin{aligned}
  & {{F}_{\theta }} \\
 & {{F}_{\varphi }} \\
\end{aligned} \right\}\exp \left( i{{\psi }_{z}}{{l}_{0}} \right)\int\limits_{-{d}/{2}\;}^{{d}/{2}\;}{d{y}'\int\limits_{a\,\,\text{or}\,\,h}^{a+b}{d{x}'\exp \left( i{{\psi }_{x}}{x}'+i{{\psi }_{y}}{y}' \right)}}},
\end{equation}
where
\begin{equation}\label{eq:5.6}
\begin{aligned}
  & {{F}_{\theta }}=-{{E}_{0x}}\cos \varphi -{{E}_{0y}}\sin \varphi +\left( {{H}_{0x}}\sin \varphi -{{H}_{0y}}\cos \varphi  \right)\cos \theta , \\
 & {{F}_{\varphi }}=\left( {{E}_{0x}}\sin \varphi -{{E}_{0y}}\cos \varphi  \right)\cos \theta +{{H}_{0x}}\cos \varphi +{{H}_{0y}}\sin \varphi ; \\
\end{aligned}\
\end{equation}
for the 1st wave the lower limit of the inner integral is $a$,
\begin{equation}\label{eq:5.7}
{{\psi }_{x}}=S-k\sin \theta \cos \varphi ,\,\,\,\,{{\psi }_{y}}={{k}_{y}}-k\sin \theta \sin \varphi ,\,\,\,\,{{\psi }_{z}}=k{{\beta }^{-1}}-k\cos \theta;
\end{equation}
for the 2nd wave the lower limit of the inner integral is $h$,
\begin{equation}\label{eq:5.8}
{{\psi }_{x}}={{\kappa }_{x}}-k\sin \theta \cos \varphi ,\,\,\,\,{{\psi }_{y}}={{k}_{y}}-k\sin \theta \sin \varphi ,\,\,\,\,{{\psi }_{z}}={{\kappa }_{z}}-k\cos \theta.
\end{equation}
Remind that expressions for ${{E}_{0x,y}}$ and ${{H}_{0x,y}}$ are given by formulas~\eqref{eq:3.10} for the 1st wave and~\eqref{eq:4.23} for the 2nd wave.

The inner integrals in~\eqref{eq:5.5} are easy to calculate. For the 1st wave, we obtain
\begin{equation}\label{eq:5.9}
\left\{ \begin{aligned}
  & {{E}_{\theta }}\left( {\vec{R}} \right) \\
 & {{E}_{\varphi }}\left( {\vec{R}} \right) \\
\end{aligned} \right\}=\frac{ik\,\exp \left( ikR \right)}{\pi R}\int\limits_{-s}^{s}{\left\{ \begin{aligned}
  & {{F}_{\theta }} \\
 & {{F}_{\varphi }} \\
\end{aligned} \right\}\frac{\sin \left( {{{\psi }_{x}}b}/{2}\; \right)}{{{\psi }_{x}}}\frac{\sin \left( {{{\psi }_{y}}d}/{2}\; \right)}{{{\psi }_{y}}}{{e}^{i{{\psi }_{x}}(a+{b}/{2}\;)\,+\,i{{\psi }_{z}}{{l}_{0}}}}d{{k}_{y}}}.
\end{equation}
For the 2nd wave, we obtain
\begin{equation}\label{eq:5.10}
\left\{ \begin{aligned}
  & {{E}_{\theta }}\left( {\vec{R}} \right) \\
 & {{E}_{\varphi }}\left( {\vec{R}} \right) \\
\end{aligned} \right\}=\frac{ik\exp \left( ikR \right)}{\pi R}\int\limits_{-s}^{s}{\left\{ \begin{aligned}
  & {{F}_{\theta }} \\
 & {{F}_{\varphi }} \\
\end{aligned} \right\}\frac{\sin \left( {{\psi }_{x}}{\left( a+b-h \right)}/{2}\; \right)}{{{\psi }_{x}}}\frac{\sin \left( {{{\psi }_{y}}d}/{2}\; \right)}{{{\psi }_{y}}}{{e}^{i{{\psi }_{x}}{\left( a+b+h \right)}/{2}\;\,+\,i{{\psi }_{z}}{{l}_{0}}}}d{{k}_{y}}}.
\end{equation}
Note that the main maximums of integrands in~\eqref{eq:5.9},~\eqref{eq:5.10} takes place with the simultaneous fulfillment of the conditions
\begin{equation}\label{eq:5.12}
{{\psi }_{x}}(\theta ,\varphi ,{{k}_{y}})=0, \quad {{\psi }_{y}}(\theta ,\varphi ,{{k}_{y}})=0.
\end{equation}
These equations determine certain dependence $\theta (\varphi )$. One can show that, for the 1st wave, eqs.~\eqref{eq:5.12} have the solution $\theta ={{\theta }_{\max 1}}=\arcsin \left( n\sin {{\theta }_{p}} \right)$ for any angle $\varphi $ (it is the round cone surface). For the 2nd wave, the equations~\eqref{eq:5.12} determine certain complex conical surface.

\section{Numerical results}

\begin{figure}
\centering
\includegraphics[width=\linewidth]{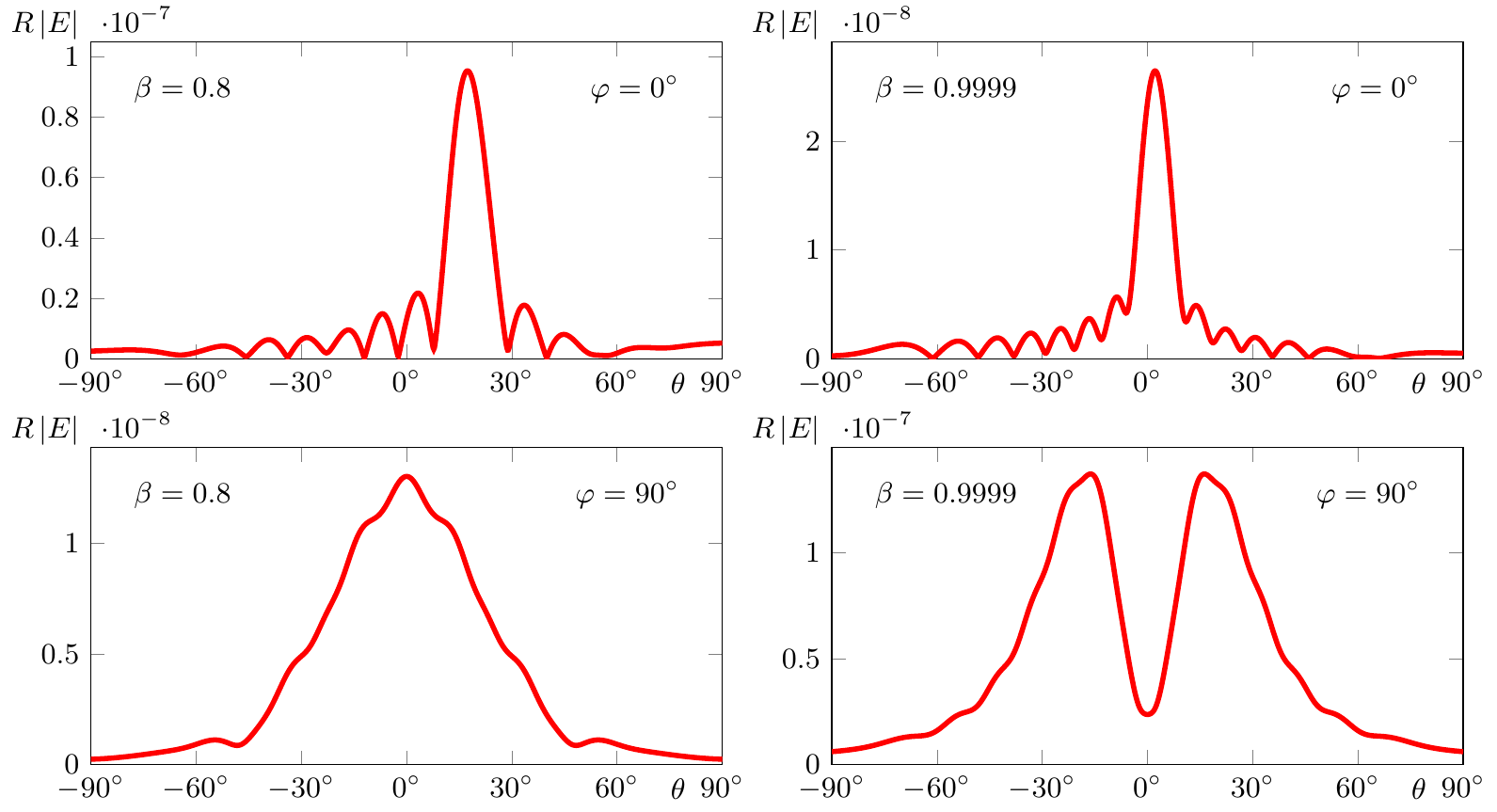}
\caption{\label{fig:4}%
  The angular distribution of the magnitude of the electric field Fourier-transform multiplied by $R$ in the Fraunhofer area (in unites $\text{V}\cdot \text{sec}$). Parameters: $q=1\,\text{nC}$, $\varepsilon =4$, $\mu =1$, $a={{k}^{-1}}$,  $b=d=50\cdot {{k}^{-1}}$,  $\alpha ={{30}^{\text{o}}}$, $\beta =0.8$ (left) and $\beta =0.9999$ (right); the oblique face is metalized. The values of $\varphi $ are shown in plots; the negative values of $\theta $  correspond to positive values of $\theta $ at replacement $\theta \to -\theta $, $\varphi \to \varphi +\pi $.
}
\end{figure}

\begin{figure}
  \centering
  \includegraphics[width=\linewidth]{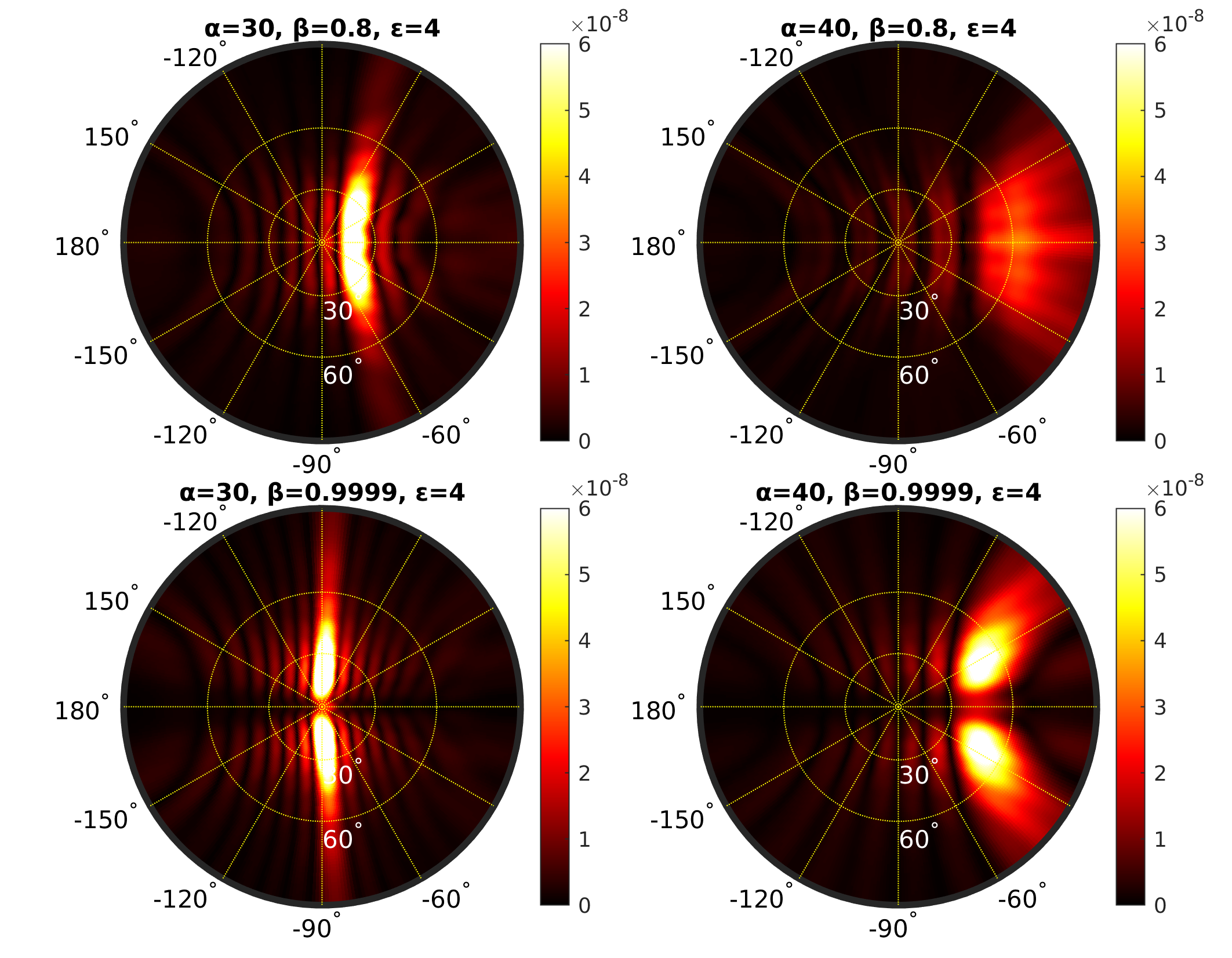}
  \caption{\label{fig:5}%
The 3-D angular distribution of the magnitude of the electric field Fourier-transform multiplied by $R$ in the Fraunhofer area  (in unites $\text{V}\cdot \text{sec}$). Parameters: $q=1\,\text{nC}$, $\varepsilon =4$, $\mu =1$, $a={{k}^{-1}}$,  $b=d=50\cdot {{k}^{-1}}$; the values of $\alpha $ and $\beta =0.8$ are shown in the plot.
  }
\end{figure}

\begin{figure}
  \centering
  \includegraphics[width=\linewidth]{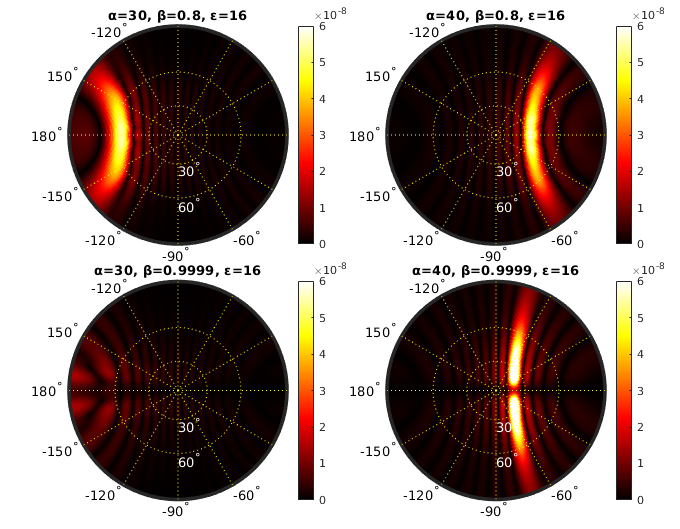}
  \caption{\label{fig:6}%
The same as in Fig.~\ref{fig:5} for $\varepsilon =16$.
  }
\end{figure}

First of all, we note that, as shown by numerical calculations, the radiation field is mainly determined by the second wave, and, as a rule, the role of the first wave is insignificant. Nevertheless, when obtaining numerical results, we took into account both of these contributions.

The results presented below refer to the case of the metallized oblique face that maximizes the radiation exiting the prism. However, it should be noted that in the absence of metallization of the oblique face, the radiation does not decrease very much. 

Figure~\ref{fig:4} shows the typical examples of the angular dependencies of the field on the angle $\theta $  for $\varphi = 0$ and $\varphi = 90^0$ (for two values of the charge velocity). The negative values of $\theta $ in this figure correspond to positive ones if we replace $\varphi $ with $\varphi +\pi $. 
So $\varphi=0$ corresponds to the ``central'' plane $y=0$, and $\varphi=90^0$ corresponds to the plane perpendicular to it. 
The vertical axis on the plots shows the value of $R\left| E \right|$ which does not depend on the distance $R$ in the Fraunhofer area. 

At any velocity, one can see the expressive maximums in both planes. 
It is interesting that, for the ultrarelativistic charge ($\beta=0.9999$), the field is minimal  at $\theta = 0$ in the plane $\varphi =90^0 $, and the maximums  take place at certain nonzero angles $\theta$ (the bottom right graph in Fig.~\ref{fig:4}). This effect is explained by the structure of the field in the key problem with the half-space, where radiation has a similar minimum in the central plane~\cite{B62}.

A more visual representation of the field structure is given by the three-dimensional color graphs, where the field magnitude  corresponds to the brightness of the color. Figure~\ref{fig:5} shows such plots in the case 
$\varepsilon=4$ for two values of the prism angle and two values of the charge velocity. Figure ~\ref{fig:6} shows similar plots in the case $\varepsilon=16$. One can see that the field structure changes essentially with variation of the problem parameters. 

Most of the plots are characterized by presence of the sharply expressed relatively small region of intense radiation. If the charge velocity is not very close to the speed of light in vacuum, there is only one such region (with the maximum field value in the central plane). Outside this region, the complex interference pattern of the field takes place, however the field magnitude is relatively small. In the case of ultrarelativistic charge velocity, as a rule, there are two regions of intense radiation which are symmetrical with respect to the central plane (as already noted, this effect is explained by the field structure in the key problem with the half-space~\cite{B62}). 

An increase in the angle of the prism $\alpha$ leads to a shift in the maximum radiation region towards the larger values of the angles $\theta$ in the region $|\varphi|<\pi/2$. An increase in the charge velocity, on the contrary, entails a shift of this region towards the smaller angles $\theta$, and, possibly, into the region $|\varphi|>pi/2$. These regularities are explained by corresponding variations in the propagation path of the second wave inside the prism. These properties are especially pronounced in the case of large values of permittivity (Fig.~\ref{fig:6}). Note that, with large deviation of the maximum radiation region from the  direction "forward" ($\theta=\phi=0$), the size of this region becomes larger, and the field strength in it decreases. This can be seen in the top right graph in Fig. ~\ref{fig:5} and, especially, in the bottom right  graph in Fig.~\ref{fig:6}, where such areas are practically absent. 

At the end of this section, we note that the radiation from the prism, as a whole, is comparable in magnitude with the radiation from the conical target having approximately the same sizes ~\cite{TGVGrig20}. However, the radiation from the prism is characterized by more complex patterns due to the absence of axial symmetry of the target. 

\section{Conclusion}

We analyzed Cherenkov radiation generated by a charge moving along one of the faces of the dielectric prism. We used the aperture method offered by us earlier, however here we have developed new version of this technique. The main difference from our previous approach consists in that we use inside the object only the expansion over the plane waves. This approach is convenient for objects having plane borders especially in the case of two or more borders on which the waves are reflected and/or refracted. The greatest advantages of this approach are manifested when calculating the field in the Fraunhofer area, since in this case the integrals over the aperture can be calculated analytically.

We analyzed the problem taking into account two waves: the 1st wave falls directly on the prism base, and the 2nd one preliminary reflects from the oblique face. The solution of the problem in the Fraunhofer area has been obtained in the form of the single integral which is very convenient for numerical calculation. The computations have shown that radiation is mainly determined by the 2nd wave. The series of plots for dependencies of the radiation field on the observation angles have been demonstrated and the main physical regularities have been noted.

\section{Acknowledgments}
This work was supported by the Russian Science Foundation (Grant No. 18-72-10137).


%

\end{document}